**When fairness is an abstraction: Equity and AI in Swedish compulsory education**

 (this manuscript was submitted to a journal and is currently under review)

Marie Utterberg Modén,

ORCID: 0000-0002-8251-8449, https://www.linkedin.com/in/marie-utterberg-7817a992/

Marisa Ponti (corresponding author),

ORCID: 0000-0003-4708-4048, https://www.linkedin.com/in/pomar/

Johan Lundin,

ORCID: 0000-0001-5547-9395, https://www.linkedin.com/in/johan-lundin-a72a782/

Martin Tallvid, https://www.linkedin.com/in/martin-tallvid-82759915/

ORCID: 0000-0002-1664-1078

Department of Applied Information Technology,

University of Gothenburg, Sweden

Box 100

405 30 Göteborg

# When fairness is an abstraction: Equity and AI in Swedish compulsory education

**Abstract**

Artificial intelligence (AI) experts often question whether AI is fair. They view fairness as a property of AI systems rather than of sociopolitical and economic systems. This paper emphasises the need to be fair of the social, political, and economic contexts within which an educational system operates and uses AI. Taking Swedish decentralised compulsory education as the context, this paper examines whether and how the use of AI envisaged by national authorities and edtech companies exacerbates unfairness. A qualitative content analysis of selected Swedish policy documents and edtech reports was conducted using the concept of relevant social groups to understand how different groups view AI's risks and benefits for fairness. Three types of groups that view efficiency as a key value of AI are identified, interpreted as economical, pedagogical and accessibility-related. By separating fairness from social justice, this paper challenges the notion of fairness as the formal equality of opportunities.

## 1. Introduction

After a series of major reforms in the early 1990s, Sweden created one of the most decentralised (from government-controlled schools to municipality-controlled schools) education sectors in the OECD (Pareliussen et al., 2019). The Swedish National Agency for Education (Skolverket) (2006) described the reforms as a general transformation of the school system, which emphasises local decision-making, competition and freedom of choice.

While it is difficult to determine how these reforms have affected Swedish schools, several evaluations of the reform effects have found that equity levels have worsened and segregation has increased (Löfstedt, 2019). This paper takes this decentralised school system as the context in which to examine the fairness implications of the use of AI. It makes us wonder whether using AI can enable or disable discrimination in education or whether it can maintain or mitigate the learning gap between those who can and cannot work with new technologies.

Whether AI is fair is not relevant, as we do not see fairness as a property of AI systems but as a property of sociopolitical and economic systems (Selbst et al., 2019). Although improperly designed systems can violate the right to education and training and perpetuate historical patterns of discrimination (European Commission, 2021), a properly designed AI system can have unintended consequences when employed by an unjust governmental or commercial regime, potentially exacerbating unfairness and power imbalances by perpetuating logics of monitoring, categorising, standardisation and social sorting (Selwyn, 2022).

Fazelpour et al. (2022) pointed out that the evaluation metrics used by developers to measure the level of fairness implemented in algorithms tend to support the goals and values of decision-makers in the same way that traditional evaluation metrics do. In education, a good example of socioeconomic class-based bias is the debacle of the algorithm used by the exam regulator Ofqual in the UK (Attwell, 2020). The data used by Ofqual showed that the algorithm downgraded the expected scores of thousands of British students, especially those with high grades from less-advantaged schools, while students from richer schools were more likely to have their scores raised (Humble, 2021). This example has shown that not including the social, political and economic contexts in which an educational system is embedded and considering fairness only in terms of allocation outputs—algorithm-based or otherwise—results in distorted evaluations and disparities. Ofqual was concerned with using a standardised method that could prevent grade inflation. Reportedly, the UK government saw this as a major problem; there were suspicions that the government did in fact support a bias in results, aiming to empower the students from elite schools to attend university, with the rest heading for a second-class vocational education and training provision (Attwell, 2020).

As the meaning of fairness is contextual, it must be understood in different educational policy contexts. Bøyum (2014) posited that educational policy depends on assumptions about fairness, whether they are made explicit or kept implicit; without a view of fairness, "one would be in the dark as to what should be done about the reproduction of social inequality through education, or whether or not anything should be done at all" (p. 858). This argument forms the basis of our study. We examine how Swedish national authorities and the edtech industry view the fairness implications of using AI in education. The decentralisation and deregulation of Sweden's compulsory education is a good context for conducting this study. Sweden used to be admired for its ability to stem inequality by using a national income-and-cost equalisation system designed to allow for the equal provision of services across the country (Pareliussen et al., 2019). With the shift towards decentralisation came choice, competition and management by objectives, resulting in the marketisation and privatisation of schools with edtech companies (Player-Koro at al., 2018). A decentralised school system makes it easier to connect with local communities, but it also challenges equal access to services and quality, as well as recruiting qualified staff, particularly in the education sector (Pareliussen et al., 2019), although an even more centralised system can have similar problems. *The aim of this study is to explore whether and how the use of AI envisaged by national authorities and edtech companies can facilitate, and perhaps exacerbate, unfairness in Swedish decentralised compulsory education.* Therefore, we address the following questions:

- How do national policymakers and edtech companies view fairness in Swedish education?
- What potential benefits are described regarding the use of AI in compulsory education, and for whom?
- What potential risks are described regarding the use of AI in compulsory education, and for whom?

Studying these questions is important because the strategies devised by national authorities influence the decisions that schools can make. Schools are embedded within the sociopolitical and economic systems of their societies, and "it is problematic to think that they can reduce achievement disparities by themselves" (Berendt et al., 2020, p. 315). We also included the edtech sector because in Swedish decentralised compulsory education, the role of this industry is important and can influence national strategies and policies.

This paper is structured as follows. Section 2 describes the existing discourse on fairness in education concerning AI. Section 3 presents the theoretical concepts, the data and the methods used in the analysis. Section 4 presents the findings before the paper concludes with a discussion and final considerations.

**2. Background**

*2.1 What do we mean by fairness in education?*

Asking whether a certain use of AI in education (e.g. algorithmic grade allocation) is fair can be given a credible response only by exploring how fairness is defined, both in a common sense and in the context in which that specific use is implemented (Clark, 2020). That is to say, asking whether something is fair should question *who* defines fairness, for whom and from what positions of power.

Bøyum (2014) observed that a common and generic view of fairness in education supports the idea of *equal educational opportunity*. His analysis of OECD policy documents on equality and education made him conclude that the OECD explicitly operates with a loose idea of equal opportunity, compatible with even merely formal equality (i.e. the absence of legal restrictions on educational access) but implicitly with a meritocratic variant of fair equality of opportunity. In this way, what fairness in education seems to mean to most people can be described: "A talented child from a poor family should have the same educational chances as a similarly talented child from a wealthy family" (Bøyum, 2014, p. 859). People need to be treated equally in their opportunities to participate in education. Maybe if we had unlimited resources and individuals with unlimited abilities, we would have achieved better equality in the conditions and opportunities afforded to people. Issues of fairness arise because we operate in a scarcity of resources, and we have conflicting views of what everyone is entitled to or how institutions should allocate scarce resources (Miller, 2021). Following Cole et al. (2022), we take fairness for whom and according to what/whose criteria as the main points to move away from an absolute and ideal view towards a view that considers the context, the different needs, circumstances and priorities of different people. Let us take the Ofqual example: When it comes to assessment, it is important to contrast grade inflation, but is it fair to dismiss teachers' predicted grades entirely in favour of standardised algorithms, letting students feel that their future has been determined by statistical modelling rather than their own abilities? This question leads us to examine the link between fairness and AI in education.

## 2.2 The focus on "algorithmic fairness"

A technical and statistical approach is prevalent within the fair AI community to address societal concerns about algorithmic bias, namely in terms of defining mathematically abstract concepts such as fairness to encode them in algorithms (Selbst et al., 2019). In fair AI, the goal of designers is to build systems that quantify bias and mitigate discrimination against certain groups (Feuerriegel et al., 2020). In practice, developers of prediction systems do not take the external reasoning and circumstances of stakeholders into account and tend to focus on a narrow set of value judgments that can be expressed as evaluation metrics (Fazelpour et al., 2022). The focus is on "algorithmic fairness" (Dignum, 2022)—on building technical solutions to potential harms rather than on interrogating the social structures, human choices and sociotechnical practices that lie behind the predicaments arising out of unfair algorithms. Unfairness and bias are social problems, and trying to solve them within the logic of computation will prove inadequate. Selbst et al. (2019) identified five "traps" that the fair AI community can fall into when designing systems; one trap central to this work is the "formalism trap", which refers to the failure to account for the meaning of concepts such as fairness, which are "procedural, contextual and contestable and cannot be resolved through mathematical formalism" (p. 61). As observed in other settings (e.g. Cole et al., 2022), we also see the gap of omitting educational policies, educational practices, schools as workplaces and teachers and students' concerns from fair AI as approached by the community of algorithmic system designers. However, all these elements should be carefully considered to translate fairness as an abstract principle into practice. It is difficult to know how to implement AI systems fairly in a specific setting without taking a contextual view of fairness into account. Thus, we consider it important to uncover the view of fairness in educational policy taken in a specific social context. To frame our analysis, we first look at how fairness and AI are linked in education. We present a brief overview of previous work addressing fair AI in education.

## 2.3     Linking fairness to AI in education

There is a risk of favouring certain individuals over others in education—as happened in the Ofqual example. Aware of the danger of encoding socioeconomic class-based bias, the European Commission (2021) has identified AI systems used in education and vocational training as high risk, as they can negatively affect fairness by exacerbating one of the best-known and most established findings of the sociology of education—the correlation between educational attainment and social background (Bøyum, 2014). A proper assessment of the beneficiaries whose goals are being fulfilled by AI, whether they are educators, students, parents, policymakers or technology companies, should be conducted to evaluate the benefits of AI (Berendt et al., 2020). AI scientists have been adapting machine learning, computer modelling and statistics techniques used in the business sector to improve decision-making in educational systems (Nistor et al., 2015). Berendt et al. (2020, p. 315) suggested that these origins raise fundamental questions about who benefits from the use of AI systems in education: learners, teachers, parents, institutions or the technology companies that create the systems. For example, AI systems designed to support students to pass an examination bear the risk of reproducing this correlation such that advantaged students "tend to have access to greater support to improve their grades, which means that grades are not a measure of ability, but a measure of what students have achieved with a high degree of supports they have access to" (Berendt et al., 2020, p. 315). Borenstein and Howard (2020) stated that the educational community, broadly defined, needs to develop the ability to recognise and engage with emerging ethical issues concerning the *concrete* implementation of AI, such as designing AI algorithms ethically, mitigating the risks of AI outcomes and improving data acquisition and other research practices. To address these issues, as Borenstein and Howard (2020) argued, vague concepts of fairness and bias, separated from the context in which these concepts apply or without understanding that people are more than just data or inputs, are not helpful. The complex ethical challenges, such as fairness, posed by AI in education cannot be addressed in a vacuum but by "starting with the root of the problem (i.e., people)" (Borenstein & Howard, 2020, p. 62).

## 3. Materials and Methods

### 3.1 Data collection

We searched the Internet and snowballed reference lists to retrieve potentially relevant documents related to AI and fairness in education in Sweden. Given the open-ended nature of our research questions, we did not perform our search using restricted search criteria to avoid missing relevant documents. Based on our initial search, we selected 12 documents, retrieved between June and September 2022, for further analysis. All the documents are in Swedish, available in the public domain and obtainable without the authors' permission. Table 1 lists these documents, which include national strategic documents and archival material, such as open accessible websites and reports provided by edtech companies. The 12 documents were produced in Sweden during the 2018–2022 period. They are meant to provide strategic guidance for governmental agencies, municipalities, and school organisers. The authors of the sampled documents display different perspectives that focus on social, technical and political issues. This diversity gives the data material coverage based on various interests and views.

**Table 1.** List of policy documents and strategic reports

| Year | Name of document | Full name and abbreviation of professional or regulatory body |
|---|---|---|
| 2022 | Branschrapport 2022. Marknadsöversikt och branschbarometer över svensk edtech – internationell tillväxt och ökat behov av livslångt lärande [Industry report 2022. Market overview and industry barometer on Swedish edtech – international growth and increased need for lifelong learning] | Swedish Edtech Industry – SEI https://media.swedishedtechindustry.se/2022/04/Swedish-Edtech-Branschrapport-2022.pdf |
| 2022 | Förslag till nationell digitaliseringsstrategi för skolväsendet 2023–2027 [Proposal for a national digitisation strategy for the school system 2023–2027] | Skolverket [The Swedish National Agency for Education] |
| 2022 | Transparens, träning och data. Myndigheters användning av AI och automatiserat beslutfattande samt kunskap om risker för diskriminering | Diskrimineringsombudsmannen – DO [Equality Ombudsman] |

| Year | Name of document | Full name and abbreviation of professional or regulatory body |
|---|---|---|
|  | [Transparency, training and data. Authorities' use of AI and automated decision-making and knowledge of risks of discrimination] |  |
| 2020 | AI och jämställdhet. En dialog om jämställdhetspolitikens genomförande i och med statliga myndigheters pågående eller kommande AI-resa [AI and gender equality. A dialogue on the implementation of gender equality policy in connection with state authorities' ongoing or upcoming AI journey] | Jämställdhetsmyndigheten - JM [Swedish Gender Equality Agency] |
| 2020 | Digitalt lärande – för att nå målen [Digital learning – to reach the goals]. | Specialpedagogiska skolmyndigheten – SPSM [National Agency for Special Needs Education and Schools] |
| 2019 | Främja den offentliga förvaltningens förmåga att använda AI [Promote the ability of public administration to use AI] | Myndigheten för digital förvaltning – DIG [Agency for Digital Government] |
| 2019 | #skolDigiplan. Nationell handlingsplan för digitalisering av skolväsendet [National action plan for digitalisation of the school system] | Sveriges kommuner och regioner – SKR [Association of Local Authorities and Regions] |
| 2018 | Automatisering av arbete. Möjligheter och utmaningar för kommuner, landsting och regioner [Automation of work. Opportunities and challenges for municipalities, county councils and regions] | Sveriges kommuner och regioner – SKR [Association of Local Authorities and Regions] |
| 2018 | Kunskapsöversikt om användningen och utvecklingen av automatiserad databehandling med algoritmer (artificiell intelligens) och stordata och diskriminering eller risker för diskriminering [Review of the use and development of automated data processing with algorithms (artificial intelligence) and big data and discrimination or risks of discrimination] | Diskrimineringsombudsmannen – DO [Equality Ombudsman] |
| 2018 | Nationell inriktning för artificiell intelligens [National approach to artificial intelligence] | Regeringen – REG [The Swedish Government] |
| 2017 | Nationell digitaliseringsstrategi för skolväsendet [National digitalisation strategy for the school system] | Regeringen – REG [The Swedish Government] |

## *3.2 Data analysis*

During our empirical analysis, we applied the theoretical concept of relevant social groups (RSGs) from the Social Construction of Technology framework (Pinch & Bijker, 1984). RSGs refer to actors that may or may not belong to the same organisation or institution, but "the key requirement" is that all members of this group share the same interpretation of an artifact (Pinch & Bijker, 1984). Actors can be members of different RSGs, but each RSG has a similar way of perceiving or interpreting a particular artefact. Different social groups have different interpretations of an artefact, and these different interpretations indicate an artefact's *interpretive flexibility* (Pinch & Bijker, 1984).

Using the 12 documents as forms of "narratives" (Bal, 2009) offering written accounts of how fairness is perceived by different actors, we identified RSGs and examined their views on fairness and the potential benefits and risks of AI application in compulsory education. We used qualitative content analysis (Hsieh & Shannon, 2005, p. 1278). Content analysis assumes that texts can provide valuable information about a particular phenomenon (Bengtsson, 2016). The unit of analysis was the individual document. NVivo 12 software was set up to code the documents.

Given the limited prior theory and evidence, we took an inductive approach to make sense of the complex data and to allow themes and categories to emerge (Eisenhardt et al., 2016). More specifically, we read all documents several times and identified the RSGs, their motives for using AI, the main problems AI can solve and the perceived benefits and risks of using AI. The process for the analysis is shown in Table 2. We conducted a manifest analysis, meaning that we remained close to the text, describing the visible, such as the words in the text, without trying to infer latent meanings (Bengtsson, 2016). Our codebook and examples of data coding are available at https://doi.org/10.6084/m9.figshare.22726883.v1. For each RSG, we created a table that displays sample quotes from the data to exemplify our codes and categories. Providing these samples can help assess whether our coding is faithful to the meanings in the data and help create clarity about each code's meaning.

| Steps of analysis | Activities |
|---|---|
| 1. Manual first round of analysis of all the reports in NVivo | • Coding each document by motives to engage with AI, perceived problems AI can address and solve, values attributed to AI in education, solutions to educational problems using AI and perceived risks of using AI<br>• Discussion of initial results within the author group<br>• Application of RSG as an analytical framework |
| 2. Manual second round of analysis of all the reports in NVivo | • Re-analysis of the findings and codes from first round of analysis |

**Table 2.** Summary of the data analysis

## 4. Results – the RSGs

We identified three main RSGs to represent the following motives: a) gain economic benefits, b) create pedagogical value and c) advocate citizen equal rights and participation. We now describe each group and their prioritised motives. The entries corresponding to the in-text citations are listed in Table 1.

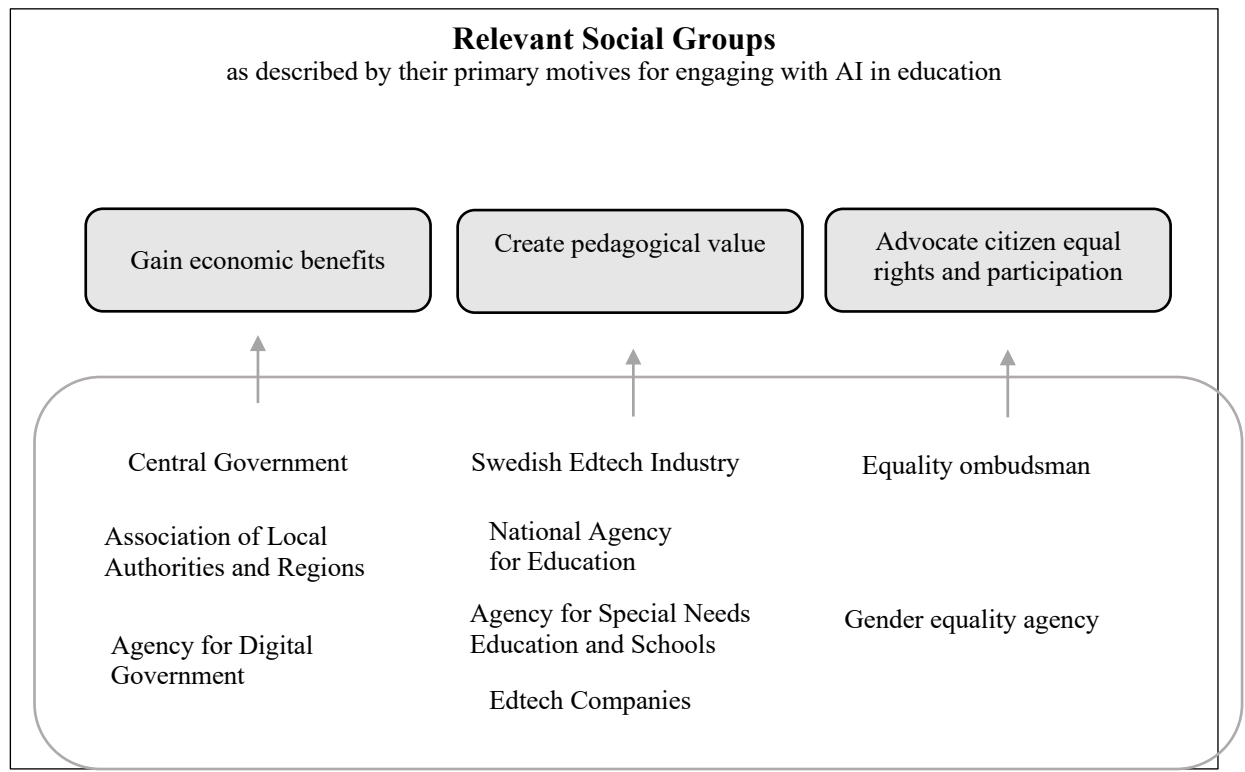

Figure 1. Identified RSGs in the analysed documents

### 4.1 RSG 1: Gain economic benefits

This RSG includes three actors: the Swedish Government (REG), the Swedish Association of Local Authorities and Regions (SKR) and the Agency for Digital Government (DIG). Its core motive lies in investing in AI to harness the economic development opportunities that AI can offer, with the aim of strengthening the quality of Sweden's public welfare and the country's international competitiveness in research and development (DIG, 2019; REG, 2018; SKR, 2018). One reason underlying this motive is that existing levels of social spending on the provision of collective goods and services—including education—are too expensive and difficult to maintain. In Sweden, the population has increased, but the percentage of people of working age has declined concurrently (SKR, 2018). The focus is on using AI to make the public sector more *effective and relevant* (REG, 2018) for citizens, for example, by partially automating jobs (DIG, 2019; SKR, 2018). This RSG makes cursory mention of the educational sector and its specific needs, suggesting that AI could improve access to more equal, high-quality education by adjusting to students' circumstances, despite declining financial conditions and a shortage of teachers (DIG, 2019). Better education has a very high economic value in society, according to the group actors. A key advantage of AI applications is their ability to provide individualised education and evaluation for the individual student and for the teacher, who is given better tools, but also to improve education in general by reducing the likelihood of students' needs not being met. In this respect, it is stated that in education, AI is used, among other things, in so-called adaptive learning. Intelligent software programs learn how an individual learns and adapt training content and feedback to that individual. This function is seen as a support for teachers, freeing up valuable time for them. The program helps teachers gain insight into where each student is and how s/he develops in his/her learning, adapts tasks and assesses the student's knowledge (SKR, 2018, p. 21). As an example of an efficiency perspective, the government suggests that AI can help increase access to high-quality education based on students' needs, despite declining country economic conditions and teacher shortages (DIG, 2019).

Among the risks associated with AI applications, this RSG identifies skewed or manipulated data, lack of transparency and misuse or hostile use. These consequences can lead to discrimination, reduced trust and economic damage and influence how democracy works (REG, 2018). Additionally, rapid changes in the labour market may lead to demands for work adjustments and unemployment. A suggested solution to reduce AI risks is to ensure that enough people are skilled at developing and using AI (DIG, 2019; REG, 2018). Collaboration with businesses, the public sector, researchers, and other EU countries is another way to create new knowledge (REG, 2018). It is also important to develop rules, standards, norms and ethical principles to guide the ethical use of AI (REG, 2018, p. 10). This RSG advocates increased engagement with AI to control and manage the risks associated with it.

## 4.2 RSG 2: Create pedagogical value

This RSG includes the Swedish edtech industry (SEI), the Agency for Special Needs Education and Schools (SPSM), and Skolverket. Using AI in schools to improve processes and teaching strategies and adapt to students' needs is at the core of this group's motives. The SEI (2022) highlights AI applications in its products and the benefits of AI for data-driven education. The SPSM and Skolverket look at school quality from a broader perspective, considering student achievement, democracy and equality. The SPSM (2020) strives to make education accessible to all students, regardless of their functional abilities. According to Skolverket's proposal to revise the national digital strategy for schools, the Swedish school system should become an international leader in utilising digitisation to develop children's and students' digital skills and promote knowledge and equality using technology (Skolverket, 2022, p. 3). In its report, the agency does not explicitly link AI to education. Rather than mentioning AI, it refers to data-driven education. As schools become more digitalised, more data that can be used to inform decisions can be generated, which is a priority for national authorities (Skolverket, 2022). In contrast, edtech companies have made such a connection clear, claiming, for example, that using AI in digital mathematics teaching tools allows teachers to include all students, as they can learn at their own pace (NE, 2018). AI-based products would free teachers up from administrative burdens, allowing them more time to do pedagogical work, resulting in better motivation, skills and well-being of students (EdAider, n.d.). In the opinion of edtech companies, efficient learning and teaching is a major issue. It is difficult for teachers to allocate time to each student's individual needs, which is where AI comes in. The motive of the edtech industry rests in using AI as a smart technology that enables data-driven education (SEI, 2022). However, there are no specific suggestions on how AI should be implemented in schools or how student learning can be improved using AI. Regarding risks, Skolverket (2022) advocates for careful data management and advises schools to ensure privacy protection when handling data and to work to prevent cyberattacks and disruptions.

### 4.3 RSG 3: Advocate citizen equal rights and participation to mitigate risks

Among the members of this RSG are government agencies, such as the Swedish Gender Equality Agency (JM) and the Equality Ombudsman (DO), which monitor compliance with the Discrimination Act (Diskrimineringslagen 2008:567) to promote a society where all people have equal rights and opportunities, regardless of their sex, transgender identity, ethnicity, religion, belief, disability, sexual orientation or age. The core motive of this group lies in preventing bias and injustice when AI applications are used. Within this group, AI is defined as an umbrella term that encompasses a wide variety of expert systems and AI applications (JM, 2020, p. 10; DO, 2022, p. 17). This group is concerned with AI to analyse data and automate decision-making in areas such as social insurance and public employment. Typical AI users are administrative officers and civil servants.

Education is one of the areas in which the DO monitors compliance with the Discrimination Act (2008). It is argued that the use of automated systems can unknowingly lead to noncompliance with the act and constitute barriers to equal rights and opportunities. Despite the vague rhetoric regarding concrete risks, the document acknowledges that individuals are at risk of unfair rejection, preventing citizens from—or reducing the possibilities of—accessing the labour market or social benefits. A few examples of risks include stereotype overrepresentation in text, voice, and image recognition (JM, 2020).

Digitalisation and automation of public activities must reflect national core values: "AI, which by its nature has proven to be fast and changeable, must rest on the same foundation as the traditional Swedish administration; the state value" (JM, 2020, p. 16). Therefore, all citizens should feel confident that decisions are made effectively and legally, regardless of gender, race, religion, or other grounds for discrimination. When applying AI and automated decision-making, the group advocates for a solution that increases knowledge about equality and discrimination risks (DO, 2022; JM, 2020). AI processes need to be explained transparently so that even nonexperts can understand them. Laws and regulations should be clear about this requirement.

## 5. Discussion and limitations

We examined how fairness is viewed in Swedish education by national authorities and edtech entrepreneurs. More specifically, we strived to understand how these actors envisage the role of AI in Swedish compulsory education, as well as their perceptions of the benefits and risks for fairness associated with AI. Following Bøyum (2014), we acknowledge that educational policy depends on assumptions about fairness, whether explicitly stated or implicitly implied. To the extent possible, our discussion in the foregoing sections is based on a close textual analysis of a set of national strategic documents and edtech business reports and qualitative evidence of what is explicitly stated and what remains implicit or vague. Additionally, we build on the related literature on AI and fairness in education. Some arguments in our discussion provide grounds for future qualitative research. For example, research is needed on how AI can transform teachers' occupations and work tasks. Similarly, future work is required to gain insights into how teachers, students and school administrators make sense of the challenges of AI for fairness by contextualising its use in actual school settings.

### 5.1 The three RSGs and their conceptualisation of fairness

According to Bijker et al. (1987), actors can coexist in different RSGs depending on how they interpret the use of an artefact. The RSGs we identified are relatively homogeneous, as they consist of national authorities and agencies and edtech companies (Figure 1). The three motives we identified can exist in parallel with prioritised motives. For example, the motive for generating economic gains can also coexist with the desire to create pedagogical value in terms of "personalised learning" facilitated by ubiquitous collection of data; advocating equal rights and participation for citizens can also be seen to fulfil the economic dimension of citizen participation, such as the right to work and the right to a minimum standard of living, thus contributing to economic gains. How these motives fuel each other could be a fruitful area for future research.

Across all three RSGs, the concept of fairness is unsurprisingly conceptualised in terms of equality of access. This is a political idea stated by the Education Act: "everyone, irrespective of their geographical location and social and economic circumstances, shall have equal access to education in schools unless otherwise stated in specific provisions of this Act" (Skollagen 2010). As stated, equal access is like minimal equality of opportunity, including "formal" equality, in that access to education is not legally restricted. Thus, minimal equality of opportunity is the belief that a competent student from a poor family should be given the same educational opportunities as a similarly competent student from a wealthy family (Bøyum, 2014). Following Bøyum's (2014) considerations, we suggest that, given the growing social inequality in Sweden, the decentralisation of compulsory education and the involvement of edtech companies, the OECD's notion of formal educational equity is narrow, separated from notions of social justice and re-articulated as formal access to education and employment.

A key component of RSG theory is the rejection of linear technology development and the assertion that technology evolves under the influence of groups of people who interpret the technology's purpose differently (Bijker et al., 1987). Our results, although suggesting different motives, do not provide evidence of different interpretations of AI's role and its perceived benefits and risks for fairness. A possible reason could be the relative homogeneity of the RSGs. In our results, we see four interconnected points in the three coexisting motives.

*5.2  Efficiency: Personalised learning and redistribution of teacher time*

As clarified by the Skolverket (2022), the constitutional comments to the Educational Act specify that equal access also means that the quality of education must be such that the established goals can be achieved throughout the country. If all children and students are provided equal access to equal offers of education, there is room to adapt teaching and the organisation of education to suit their needs (i.e. personalised learning) (Skolverket, 2002, p.

8). The national authorities and edtech companies do not define precisely what personalised learning is, but they are explicit about their vision for what learning *should be like* and how AI and the massive amount of data available are expected to play a role in it. There are different meanings of personalised learning, for example, students moving at their own pace through lessons and assignments, unlike classrooms where everyone is expected to move through material together (Utterberg et al., 2021), or personalised learning driven by students' varied abilities or needs. The implicit emphasis is on the efficiency and quality of the "service" offered. Data allow AI to profile students and offer appropriate individual plans by automating tasks such as assessment, plagiarism detection, administration and feedback. This would result in teachers changing their roles, redistributing their workloads and undertaking new professional development to prepare them to teach with AI. We endorse that "what is not yet clear is how this will happen" (Miao et al., 2021, p. 18). It is a complex process, as defining new teachers' roles and competencies requires strengthening teacher training institutions and establishing appropriate capacity-building programs to prepare teachers for working with AI (Miao et al., 2021, p. 27).

*5.3    Widening social distances: Hampering formal educational equity*

The political idea of educational equity affirmed by the Educational Act is hampered by widening social distances. As a result, fairness becomes an abstract ideal "mainly serving as an expression, or allegory of our concern with social inequalities" (Blumsztajn, 2020, p. 834). The examined documents do not mention the unavoidable tension between equality and selection introduced into the decentralised school system, affected by higher local needs and/or a thinner revenue base caused by, for example, demographics, internal and external migration patterns and differences in the strength and structure of the local labour market (Pareliussen et al., 2019). There is no reference to the impact of the allocation of public money on schooling or how class segregation affects compulsory education (independent schools in privileged neighbourhoods vs. municipal schools in disadvantaged areas). According to the Swedish Ministry of Education (Edholm: Ersättning till friskolor ses över, 2022), municipal schools cost more than independent schools. The former attracts "cost-effective" children of well-educated Swedish-born parents; the latter attracts "costly" children of low-skilled, immigrant parents (Böhlmark et al., 2016). Disadvantaged schools have teachers with less experience than schools with more affluent students. Competence development can also be a problem in municipal schools. As independent schools enjoy more financial resources, it is expected that new technologies and AI will become more common in such schools, and teachers will have more opportunities for building new competencies to teach effectively with AI, understand AI's potential benefits and risks and ensure its fair use. "The skills gap between those who can and cannot work with the new technologies will continue to grow" (Miao et al., 2021, p. 23). Decentralisation allows more proximity to local communities, but it entails challenges to equality of access and quality of services across the country and recruitment of qualified staff, notably education and health personnel (Pareliussen et al., 2019). Higher local needs and/or a thinner revenue base caused by demographics, internal and external migration patterns and differences in strength and structure of local labour markets are met by a national income-and-cost equalisation system designed to allow an equal provision of services across the country. Arguably, the effects of school decentralisation do not support the idea that "quality education should be such that the established goals can be achieved all over the country", as stated in the constitutional comments to the Educational Act. As noted by Blumsztajn (2020), the idea of educational equity forces the educational system to legitimise social inequalities and unfairness, as this means treating students equally while they do not have the same social and economic opportunities. Despite general concerns regarding the use of AI in terms of bias, lack of transparency and discrimination, we contend that Swedish national digital strategies do not provide an accurate assessment of the challenges connected to AI in education, including fairness. These challenges can be exacerbated by the decentralisation of the school system due to socioeconomic factors.

*5.4    AI as a universalist technology: The specificity of education not considered*

Across the three RSGs, the focus is on public services. Far less attention is paid to the education sector and its specific needs. Education is a specific area for the application of AI, with a different focus from other areas (Molenaar, 2022). This lack of attention to the specific implications of AI in education goes along with the conception of AI as technology with a "universalist or unbounded horizon" (Cole et al., 2022, p. 7). This universalist view reflects the design of AI systems based on abstracting away the domain-specific aspects of a problem so that the systems can be used across many domains (Selbst et al., 2019). Another element contributing to the universalist view can be seen in AI scientists having adapted machine learning, computer modelling and statistics techniques used in the business sector to improve decision-making in educational systems (Nistor et al., 2015). Redrawing the boundaries of abstraction used by AI scientists to include educators and schools could help consider *reasoning and facts that are used,* for example, by school administrators to make decisions about the allocation of scarce resources, such as school support and teacher competence development, or by teachers to adapt their curricula to their classes. To ensure fairer educational outcomes, fairness needs to be understood not only as an abstract concept but also as a contextually sensitive norm that is translated into concrete sociotechnical design requirements that support this value.

*5.5   Collaboration between the educational sector and edtech businesses*

The field seems vulnerable to "hype cycles, short-termism, and forgotten histories" (Williamson & Enyon, 2020, p. 233). As a result of these ahistorical approaches, businesses can play a central role in shaping and investing in educational futures (Williamson & Enyon, 2020). It is predicted that by 2027, revenues from AI in education will reach $20 billion, which is an exponential growth in edtech since 2020, when revenues amounted to $1 billion (Brusoni, 2022). The decentralisation of the school system has led to the involvement of edtech companies in the Swedish system. However, collaboration between schools and edtech, if based on accountability and transparency of work, might be crucial for the development of AI and the collection of data, with positive consequences for fairness. Schools should not be seen as mere users of AI but should be encouraged to take a more active role in the co-creation of AI solutions by providing their ideas and perspectives to protect students and improve their learning and encouraging the creative use of technology through experimentation and pilot projects. Involving schools and educators in close collaboration with edtech could help not to succumb to Ulysses' sirens, that is, the hype that AI will solve all the pressing educational problems, based on misunderstanding current AI technical possibilities, lack of knowledge about state-of-the-art AI in education and narrow views on the role of education in society (Holmes & Tuomi, 2022). To contrast this hype with a sober examination of the advantages and disadvantages of using AI, teachers and school administrators must engage their social contexts in the development of these systems with edtech to reclaim a central role in shaping and investing in educational futures.

*5.6   Limitations*

This study is not without limitations and points to further research desiderata. First, the practical and funding circumstances under which we conducted this study allowed for a single coding of the documents. While we realise that interjudging reliability is often perceived as the standard measure of research quality to validate a coding scheme, some scholars argue that a single researcher conducting all the coding can be sufficient in studies where being familiar with the research topic is critical for the quality of the data collected (e.g., Morse & Richards, 2002). To improve the relevance of our interpretations of the data, we discussed the results of the analysis within our research group. The first and third authors were familiar with the Swedish educational system and had already investigated the use of AI-based tools in compulsory education. Furthermore, the first and fourth authors are trained teachers. We also discussed our preliminary results at a workshop on "(Un)fairness of Artificial Intelligence" organised at the University of Amsterdam in October 2022.

Second, we rely only on documents; this does not allow us to access deep knowledge or details. Additional methods, such as interviews, could have provided more insight into the RSGs' perceptions of the fairness implications of using AI in schools. Third, the findings for the Swedish case may well travel to similar countries for decentralised and deregulated school systems. However, corroboration of these results in other contexts will require further research.

## 6. Conclusion

Despite its limitations, this study presents important empirical insights into current academic and policy debates about how the use of AI envisaged by national authorities and edtech companies facilitates and potentially exacerbates unfairness in decentralised Swedish compulsory education. This exploratory study does not contribute to evidence of plausible causal relationships, given the complexity of the school system and the difficulty of isolating causal effects.

Our analysis indicates that within decentralised compulsory education, the political idea of fairness as educational equity is narrow, separated from the notion of social justice and articulated as formal access to education and employment. Neighbourhood segregation and school choice, largely made by students from privileged backgrounds, lead to school segregation. Where students live and where they attend school from the age of seven have a significant impact on their performance and their ability to benefit from the valuable educational use of AI by experienced and competent teachers. We assert that fairness is not a property of AI systems but a political value. However, we do not ignore the importance of the design and development of those systems, which can certainly threaten fundamental rights and democracy, but we take an ecosystemic view, stressing the ability of the sociopolitical and economic contexts to be fair in which an educational system operates and which will use AI.

For this reason, we have not considered several existing fairness criteria and principles for trustworthy AI, as they understandably focus on the design of technical systems. AI can positively impact education and students' lives if we, as humans, strive to make it happen. People, and particularly teachers, not AI systems, are the ones who can strive for improvement in education. In educational systems where AI is used, teachers should be engaged in developing a substantive and nuanced understanding of what AI is and what it can do. Filling the existing educational gaps in decentralised compulsory education, especially with regards to demographics and rural areas, can prevent AI from exacerbating unfairness, mitigating bias and social injustice.

**Acknowledgements**

This work was supported by the [Funding Agency] under Grant [number xxxx].

**Declaration of interest statement**

The authors report that there are no competing interests to declare.

**Appendices (as appropriate)**

Authors (2023). Codebook and examples of data coding using NVivo. Figshare. Dataset. https://doi.org/10.6084/m9.figshare.22726883.v1